# Performance Improvement of Heterogeneous Wireless Networks using Modified Newton Method

R.Shyam Sundar[1] and S.Nanda Kumar[2]

[1]School of Electronics Engineering, VIT University, *Vellore-632014, Tamilnadu, India.* `haishyamsundar@gmail.com`

[2]School of Electronics Engineering, VIT University, *Vellore-632014, Tamilnadu, India.* `snandakumar@vit.ac.in`

## ABSTRACT

*The heterogeneous wireless networks where coexistence of different Radio access technology (RAT) are widely deployed for various services and support various traffic demand, channel allocation. Under heterogeneous wireless networks, a user can send data through a single or multi RATs simultaneous. The objective of this paper is to choose the optimal bandwidth for the services and power allocation to that bandwidth. The proposed distributed joint allocation algorithm using modified Newton method is adopted to maximize the total system capacity. We validate the performance of the proposed algorithm through numerical results.*

## KEYWORDS

*Heterogeneous Wireless Networks, Multi-Radio Access (MRA), Radio Access Technology (RAT), Joint Allocation, Radio Resources Management (RRM).*

## 1. INTRODUCTION

The development and proliferation of wireless and mobile technologies have revolutionized the world of communications. Such technologies as Bluetooth and ultra wideband (UWB) radios for personal areas, wireless local area networks (WLANs) for local areas, Worldwide Interoperability for Microwave Access (WiMAX) for metropolitan areas, 3G cellular networks for wide areas, and 3GPP Long Term Evolution(LTE) [1].The integration of different technologies is known as heterogeneous wireless network. The heterogeneous wireless network concept for beyond 3G systems is intended to propose a flexible and open architecture for a large variety of different Radio Access Technologies(RAT), applications and services with different QoS demands, as well as different protocol stacks. The concept of heterogeneous wireless network is always being connected in always best connected (ABC) fashion [5]. The realization of the ABC concept where several RATs coexist calls for the introduction of new radio resource management (RRM) strategies operating from a common perspective that takes into account the overall amount of resources in the available RATs, and therefore are referred to as CRRM (Common Radio Resource Management) algorithms [16].

The multi radio access (MRA) enables networks utilizing several access techniques to communicate [6]. MRA can be accomplished in two ways: switched MRA can connect one Radio

DOI : 10.5121/ijsea.2012.3307        79

Access Technology at a time and parallel MRA can connect over multiple radio access technology simultaneously. The network optimization has to increases the total system capacity and improves the system connectivity as well as system efficiency [7]. For the transmission of the data, system has to choose the optimal bandwidth and power allocation for the optimal solution

## 2. RELATED WORKS

Some promising MRA concept for improving the total system capacity may be found in literature [2-5] related with Ambient Networks (AN). In [2] author Feasibility Studies on the MRA architecture is the efficient utilization of the multi-radio resources by means of effective radio access selection mechanisms. The Multi-Radio Access (MRA) architecture consists of Multi-Radio Resource Management (MRRM) and Generic Link Layer (GLL) functionality. The MRRM consists of Radio Access (RA) coordination and network-complementing RRM functions. RA coordination functions are load/congestion control, RA advertisement, RA discovery, RA selection and Radio Resource Monitoring. The RA advertisement function serves to display a networks ability to communicate and cooperate with other networks. This can be detected by the RA discovery function, which identifies possible RAs including multi-hop routes. Admission and bearer selection are then granted by the RA selection function [7]. The Radio resource Monitoring function provides uniform data (e.g., different network load measures) as input to other MRRM functions. In author address the dynamic channel allocation with QoS support and the comparisons between the cooperative and non cooperative resource allocation. In addition, [8] shows the resources allocation, congestion control and scheduling algorithm for multi RAT and optimization problem in the selection of optimal solution of resources such as RAT, bandwidth, power for the parallel MRA [10]. We investigate the optimal resources allocation issues to increases the total system capacity of the parallel MRA.

The rest of the letter is organized as follows. Section 3 begins with introducing the parallel MRA and system model. In section 4 algorithms for optimal solution is provided and section 5 simulation results followed by conclusion in section 6.

## 3. SYSTEM MODEL

As shown in Fig. 1. We consider MRA system model based on heterogeneous wireless networks. It consist of many subsystem (i.e. radio interfaces) available for each MMT (Multi mode, multi band user terminal) by implementing cognitive radio (CR) over software defined radio (SDR) technology [9]. The presence of multi radio access techniques (RAT) is able to improve the total system performance is known as RAT diversity.

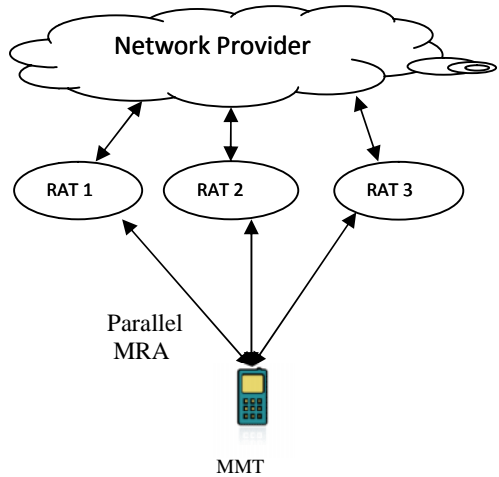

Figure 1. Parallel MRA System.

Each MMT has capable of transmitting data through different RATs and operating bandwidth. Each subsystem has its own operating frequencies and its bandwidths. In MRA system data can be transmitting through switched or parallel MRA [1].

The MRA system transmit data in two processes namely a) user scheduling and b) radio access scheduling. The user scheduling process refers to scheduling of user and its operation is steered by user satisfaction. The user scheduling is divided into two groups. a) Fast scheduling and b) slow scheduling. The radio access schedulers that select the RAs via which the scheduled data unit (MAC or IP PDU) are going to be transmit [3]. Regardless of the number of interfaces, perfect synchronization among subsystem and MMTs is assumed for necessary signal exchange and data transmission.

## 4. PROBLEM FORMULATION AND OPTIMALITY CONDITIONS

Heterogeneous wireless network is shown in figure. There are *L* MMT and *k* RATs. Based up on demand for spectrum, RATs *q* provides bandwidth     units to the *L* MMTs. After allocating bandwidth, each MMT experiences different channel gain on each bandwidth. The channel gain to noise ratio for MMT p and RAT q can be indicated by

$$c_{pq} = \frac{|H_{pq}|^2}{N_{pq}} \qquad (1)$$

Whereas $H_{p\,q}$ is channel transfer function and $N_{pq}$ is total noise power spectral density. Let $b_{pq}$ be the bandwidth obtained by MMT *p* from RAT *q*. each MMT *q* transmits his data over bandwidth $b_{pq}$ at rate    . therefore from Shannon capacity formula for Gaussian channel

The achievable data rate ($d_p$) of MMT *p* as Whereas K is the total number of RATs an MMT $p$. $p_{pq}$ be the transmission power of MMT $p$ to RAT $q$, $q$ ($0 \leq \beta_q \leq 1$) represents the efficiency which can be guaranteed by RAT $q$ to an MMT.

The maximization problem for MRA allocation can be formulated as

$$(P) \max R(b,p) = \max \sum_{p=1}^{L} d_p$$

$$= \max \sum_{p=1}^{L} \sum_{q=1}^{K} \beta_q b_{pq} \log\left(1 + \frac{c_{pq} p_{pq}}{b_{pq}}\right) \qquad (3)$$

Subject to

$$\sum_{p=1}^{L} b_{pq} \leq B_q, \quad \forall q \qquad (4)$$

$$\sum_{q=1}^{K} p_{pq} \leq P_p, \quad \forall p \qquad (5)$$

$$b_{pq}, p_{pq} \geq 0 \qquad (6)$$

Where $L$ is the total numbers of MMTs, $B_q$ is the total system bandwidth of RAT q and $P_q$ is the maximum power of MMT.

For the optimal solution of problem (P), the Lagrangian can be formulated as

$$L(b_{pq}, p_{pq}, \lambda_q, \mu_p) = \sum_{p=1}^{L} \sum_{q=1}^{K} \beta_q b_{pq} \log\left(1 + \frac{c_{pq} p_{pq}}{b_{pq}}\right)$$

$$+ \sum_{q=1}^{k} \lambda_q \left[ B_q - \sum_{p=1}^{L} b_{pq} \right] + \sum_{p=1}^{L} \mu_p \left[ P_p - \sum_{q=1}^{K} p_{pq} \right] \qquad (7)$$

Where $\lambda_q$ and $\mu_p$ are shadow prices with non negative Lagrange multipliers. Based on the karush-kuhn-Tucker (KKT) condition for the optimization problems.

$$\frac{\partial L}{\partial \beta_{pq}} = \beta_q \log\left(1 + \frac{c_{pq} p_{pq}}{b_{pq}}\right) + \beta_q b_{pq} \left(\frac{b_{pq}}{b_{pq} + c_{pq} p_{pq}}\right) \left(\frac{b_{pq} - (b_{pq} + c_{pq} p_{pq})}{b_{pq}^2}\right) - \lambda_q$$
$$\leq 0 \quad (8)$$

$$= \beta_q \log\left(1 + \frac{c_{pq} p_{pq}}{b_{pq}}\right) + \beta_q b_{pq} \left(\frac{b_{pq}}{b_{pq} + c_{pq} p_{pq}}\right) \left(\frac{-c_{pq} p_{pq}}{b_{pq}^2}\right) - \lambda_q \leq 0$$

$$= \beta_q \log\left(1 + \frac{c_{pq} p_{pq}}{b_{pq}}\right) - \beta_q \left(\frac{c_{pq} p_{pq}}{b_{pq} + c_{pq} p_{pq}}\right) - \lambda_q \leq 0 \qquad (9)$$

$$\frac{\partial L}{\partial p_{pq}} = \beta_q b_{pq} \left(\frac{b_{pq}}{b_{pq} + c_{pq} p_{pq}}\right) \left(\frac{c_{pq}}{b_{pq}}\right) - \mu_q \leq 0$$

$$\frac{\partial L}{\partial p_{pq}} = \frac{\beta_q c_{pq} b_{pq}}{b_{pq} + c_{pq} p_{pq}} - \mu_q \leq 0 \tag{10}$$

$$b_{pq} \left[ \beta_q \log\left(1 + \frac{c_{pq} p_{pq}}{b_{pq}}\right) - \frac{\beta_q c_{pq} p_{pq}}{b_{pq} + c_{pq} p_{pq}} - \lambda_q \right] = 0$$

$$p_{pq} \left( \frac{\beta_q c_{pq} b_{pq}}{b_{pq} + c_{pq} p_{pq}} - \mu_q \right) = 0 \tag{11}$$

$$\lambda_q \left[ B_q - \sum_{p=1}^{L} b_{pq} \right] = 0 \tag{12}$$

$$\mu_p \left[ P_i - \sum_{q=1}^{K} p_{pq} \right] = 0 \tag{13}$$

Using (10) and (11) the relationship between BW and power allocation can be obtained

$$\frac{\beta_q c_{pq} b_{pq}}{b_{pq} + c_{pq} p_{pq}} - \mu_q \leq 0$$

$$\frac{\beta_q c_{pq} b_{pq}}{b_{pq} + c_{pq} p_{pq}} = \mu_q$$

$$\beta_q c_{pq} b_{pq} = \mu_q (b_{pq} + c_{pq} p_{pq})$$

$$\mu_q b_{pq} + \mu_q c_{pq} p_{pq} = \beta_q c_{pq} b_{pq}$$

$$\mu_q c_{pq} p_{pq} = \beta_q c_{pq} b_{pq} - \mu_q b_{pq}$$

$$p_{pq} = \frac{\beta_q c_{pq} b_{pq} - \mu_q b_{pq}}{\mu_q c_{pq}}$$

$$p_{pq} = b_{pq} \left[ \frac{\beta_q}{\mu_p} - \frac{1}{c_{pq}} \right]^+ \tag{14}$$

Where $[z]^+ = max\{z, 0\}$ from this we can get optimal $b_{pq}$ and $p_{pq}$ value. The proposed technique Modified Newton method can be applied to $b_{pq}$ because it is global convergence toward a local maximum than other algorithms such as steepest descent method. It's satisfy all properties such as descent property, quadratic termination property, global convergent, order of convergence i.e. p=2 [18].
Now lets us take function

$$f(b_{pq}^n) = \beta_q \log\left(1 + \frac{c_{pq}p_{pq}^n}{b_{pq}^n}\right) - \beta_q\left(\frac{c_{pq}p_{pq}^n}{b_{pq}^n + c_{pq}p_{pq}^n}\right) - \lambda_q^n \quad (15)$$

$$f'(b_{pq}^n) = \frac{c_{pq}p_{pq}^n}{b_{pq}^n + c_{pq}p_{pq}^n}\left(\frac{\beta_q}{b_{pq}^n + c_{pq}p_{pq}^n} - \frac{1}{b_{pq}^n}\right) \quad (16)$$

Whereas superscript *n* represent the $n^{th}$ iteration. And the optimal bandwidth value can be obtained by Newton method and modified Newton method respectively.

$$b_{pq}^{n+1} = b_{pq}^n - \frac{f(b_{pq}^n)}{f'(b_{pq}^n)} \quad (17)$$

$$b_{pq}^n - \frac{f(b_{pq}^n)}{f'(b_{00}^n)} \quad (18)$$

After calculating optimal bandwidth, power can be calculate using eqn. taking the derivative with respect $p_{pq}$ give the KKT condition corresponding to the usual waterfilling level $(n_p)$ of each MMT p can be represent as

$$\begin{cases} \frac{p_{pq}^n}{b_{pq}^n} + \frac{1}{c_{pq}} = n_p, & \text{if } p_{pq}^n > 0 \\ \frac{1}{c_{pq}} \geq n_p & \text{if } p_{pq}^n = 0 \end{cases} \quad (19)$$

Let the continuously differentiable dual function for updating $\mu_p^n$ and $\lambda_q^n$ value for optimal solution.

$$D(\lambda_q, \mu_p) = \max_{b,p} L(b_{pq}, p_{pq}, \lambda_q, \mu_p) \quad (20)$$

Update the $\mu_p^{n+1}$ value for power allocation is given by

$$\mu_p^{n+1} = \left|\mu_p^n - \xi\frac{\partial D(\lambda_q^n, \mu_p^n)}{\partial \mu_p^n}\right|^+$$

$$= \left|\mu_p^n + \xi\left[\sum_{q=1}^{K} p_{pq} - P_p\right]\right|^+ \quad (21)$$

Whereas $\xi$ is a constant step size ($\xi > 0$). For update the $\lambda_q^{n+1}$ value for bandwidth allocation is given by

$$\lambda_q^{n+1} = \left|\lambda_q^n - \xi\frac{\partial D(\lambda_q^n, \mu_p^n)}{\partial \mu_q^n}\right|^+$$

$$= \left|\lambda_q^n + \xi\left[\sum_{p=1}^{L} B_{pq} - B_q\right]\right|^+ \quad (22)$$

From the iteration (14)-(22) we get the optimal solution for maximize total system capacity.

## 5. SIMULATION RESULTS

To evaluate the performance of joint resource allocation technique for maximize the total system capacity. We consider two RATs, bandwidth of 5MHz and 20MHz with same efficiency (i.e. $\beta_q = 1$ for q = 1, 2) and distance between the access point is 200metres. Total power consumed by each MMT is 20mW.

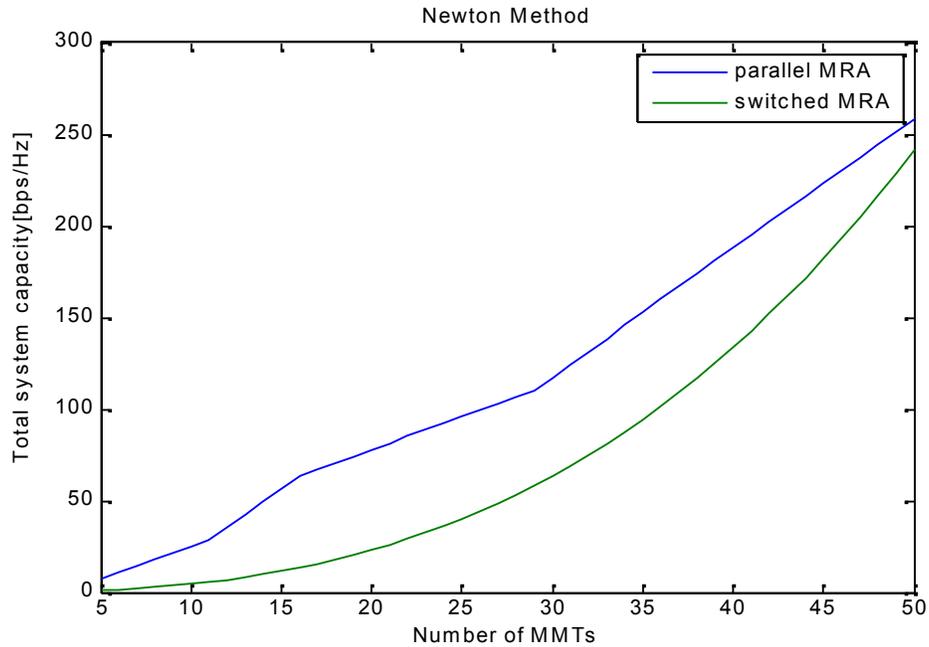

Figure 2. The comparison of parallel and switched MRA with number of MMTS using Newton method

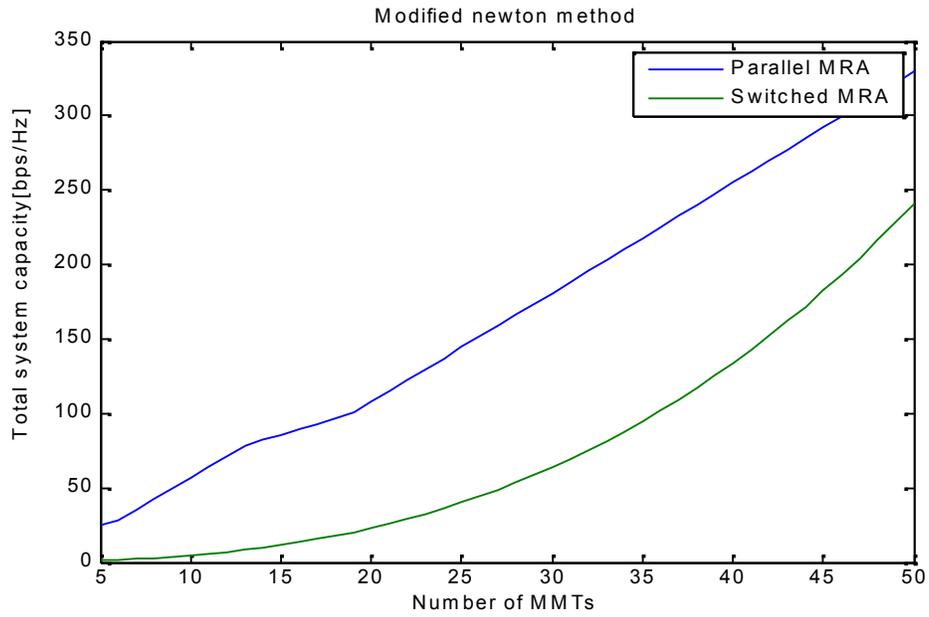

Figure 3. The comparison of parallel and switched MRA with number of MMTs using modified Newton method

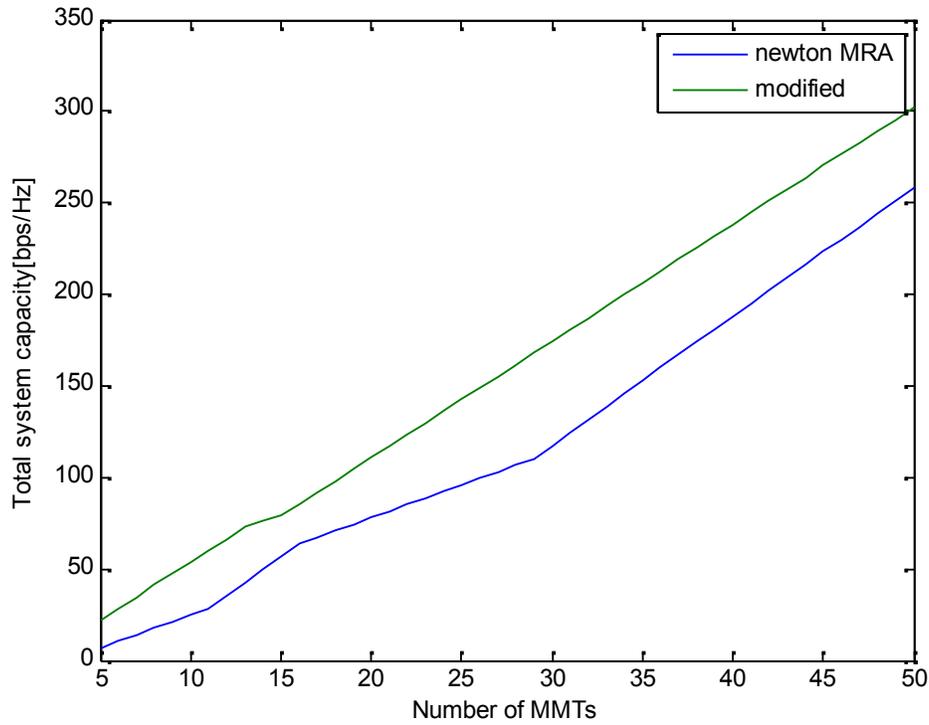

Figure 4. Comparison of parallel MRA with number of MMTs.

| TOTAL OF MMTs | NEWTON METHOD<br><br>TOTAL SYSTEM CAPACITY<br>[bps/hz] | MODIFIED NEWTON METHOD<br>TOTAL SYSTEM CAPACITY<br>[bps/hz] |
|---|---|---|
| 5 | 7.09 | 21.99 |
| 10 | 24.79 | 53.96 |
| 15 | 56.56 | 79.05 |
| 20 | 77.95 | 110.76 |
| 25 | 88.69 | 129.89 |
| 30 | 106.45 | 161.67 |
| 35 | 138.63 | 193.55 |
| 40 | 174.00 | 225.59 |
| 45 | 209.11 | 257.43 |
| 50 | 258.03 | 302.00 |

Table 1. Comparisons of parallel MRA

The Figure 1 shows the comparison of parallel and switched MRA with number of MMTs using Newton method. From that we concluded the total system capacity of parallel MRA is increased compared to switched MRA because parallel MRA can connect over multiple radio access technology simultaneously, whereas switched MRA can connect one radio access technology at a time. The coexistence of multi RATs enhance the total system throughput.

Figure 2. The total system capacity increases compared to the Newton method. The modified Newton converges faster towards a local maximum because Newton method is lack of global convergence property. It's satisfying all properties such as descent property, quadratic property, global convergence and order of convergence. From Figure 3. The total system capacity at the 25 number of MMT for Newton and modified method is 92.28 and 136.20 respectively. In modified Newton total system capacity increases up to 67% compared to the existing Newton method.

In Figure 4. Apply the algorithm to find optimal solution when the modified Newton method is applied it can be seen that MMT 1 chose both RAT 1 and RAT 2. And chose only one RAT after iterative calculative for maximize the system capacity. The bandwidth utilization is minimum when compares to the existing method.Figure 5. Shows the shadow prices for the allocated bandwidth. The shadow prices increases exponential for spectrum allocated bandwidth for each MMT. Therefore in parallel MRA each MMT accessing of different RAT depend up on bandwidth and power constraints.

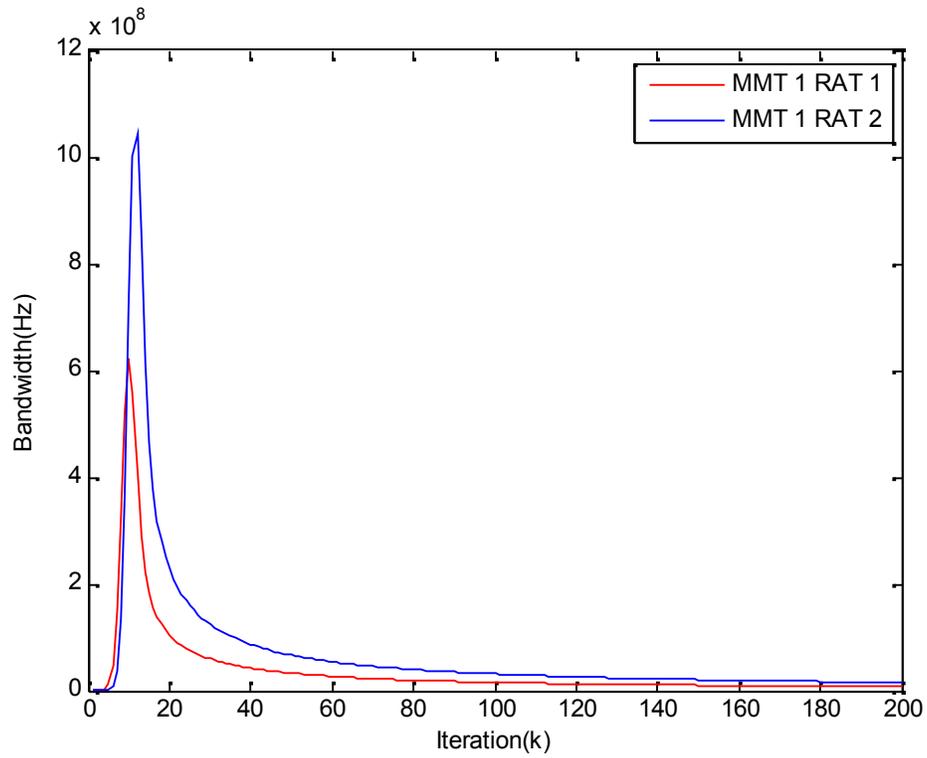

Figure 5. An illustration to find the optimal solution when the modified Newton method is applied

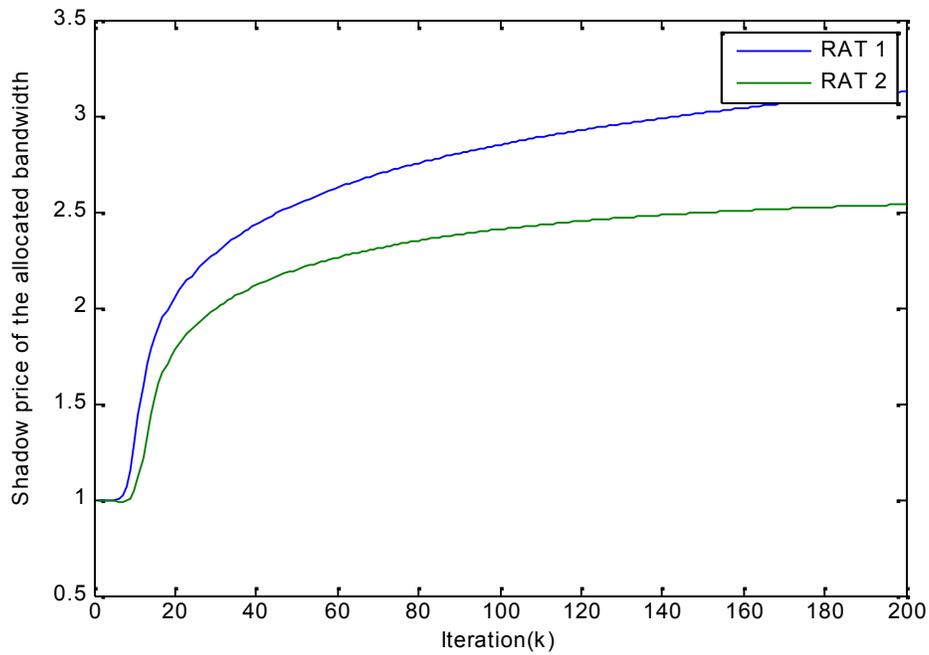

Figure 6. An illustration of the corresponding shadow price when the modified Newton method is applied

## 6. CONCLUSION

Total system capacity is improved by distributed joint allocation algorithm using modified Newton method. Evaluation results show that the modified Newton method performance well as compared to Newton method in MRA system. The parallel MRA is superior to the switched MRA method because of multi-RAT diversity, utilization of optimal bandwidth and power allocation. The conjugate gradient method with reduced complexity and better order of convergence remains for future work.

## APPENDIX

Consider the unconstraint optimization problem to get a global min point $\text{Min}_{x \epsilon R^n} f(x)$
Let us assume that the function $f$ is dual continuously differentiable

### A. Descent Property

The objective function () values should decrease as we proceed through the sequence $\{x^k\}$, i.e. $f(x^{(K+1)}) < f(x^k)$ for all k.

### B. Quadratic Termination Property

The objective function is said to be quadratic termination property if the minimum of positive definite quadratic form in n variables is reached in at most n iterations.

### C. Globally Convergent

The objective function is said to be globally convergent if starting from any point $x$   $R^n$, the sequence always converges to a point of the solution set .

### D. Order of Convergences

Let the sequence $\{x^k\}$ convergence to a point $\bar{x}$ and lets $x^{(k)}$   $\bar{x}$ for sufficient large k. the quantity $\|x^{(k)} - \bar{x}\|$ is called the error of the k$^{\text{th}}$ iterate $x^{(k)}$. Suppose that there exists $p$ and $0 < a <$   such that

$$\lim_k \frac{\|x^{(k+1)} - \bar{x}\|}{\|x^{(k)} - \bar{x}\|^p} = a \ (0 < a < \ )$$

Then $p$ is called the order of convergence of the sequence $\{x^k\}$. Thus $\|x^{(k)} - \bar{x}\|$ = a$\|x^{(k)} - \bar{x}\|^p$ asymptotically. If p=1, the sequence $\{x^k\}$ is said to have linear convergence rate. Anf for $p$=2, it is said to have quadratic convergence rate. In case $p$=1 but a=0, then the sequence $\{x^k\}$ is said to have super liner convergence rate.

The order of convergence tells us how the tail of sequence $\{x^k\}$ behaves. Larger values of $p$ will converge faster. Most of the algorithm do very well for first few iteration but become very slow near the optimal solution. But if $p$ is large then there will be significant improvement in the objective function value even near the actual solution.